\documentclass[prb,aps,twocolumn]{revtex4}
\usepackage{amsfonts}
\usepackage{amssymb,latexsym}
\usepackage{amsmath}
\usepackage{epsfig}
\usepackage{graphics}

\begin{document}

\title{Quantum versus classical quenches and the broadening of wave packets}

\author {K. Sch\"onhammer}
\affiliation{Institut f\"ur Theoretische Physik, Universit\"at
  G\"ottingen, Friedrich-Hund-Platz 1, D-37077 G\"ottingen}

\date{\today}

\begin{abstract}

  The time dependence of  one-dimensional quantum mechanical
  probability densities
  is presented when the potential in which a particle moves is
  suddenly changed, called a quench.
  Quantum quenches are  mainly addressed but a comparison with
  results for the dynamics in the framework of classical
  statistical mechanics is useful. 
  Analytical results are presented
  when the initial and final potentials are harmonic
  oscillators. When the final potential vanishes the problem reduces
  to the broadening of wave packets.
  A simple introduction to the concept of the Wigner function is
  presented which allows a better understanding of the dynamics of
  general wave packets. It is pointed out how
  special the broadening of Gaussian wave packets is, the only example
  usually presented in quantum mechanics textbooks.

 \end{abstract}
 \maketitle

\section{Introduction}

A quantum quench is a sudden change of the
Hamilton operator of a system. The  initial operator
 $\hat H_i$ is changed to the final one $\hat H_f$.
The study of the dynamics
 after the quench is an active field of research
 in  quantum many-body systems \cite{CC} as this presents
 an interesting example of non-equilibrium physics.

In this paper, we present one simple example of such a quench.
Our discussion for a  single particle in one dimension
complements standard  courses as
only a  few time-dependent problems are  discussed in
most  quantum mechanics textbooks.

As the initial state we take the ground state $|\psi_0\rangle$
of the initial Hamiltonian
$\hat H_i=\hat T+\hat V_i$, where $\hat T=\hat p^2/2m$ is
the operator of the kinetic
energy and $ V_i(x)$ is the initial attractive potential.
At $t=0$
the change to $\hat H_f=\hat T+\hat V_f$,  is performed.
The time dependence can be described in the Schr\"odinger or the
Heisenberg picture. Which is more appropriate depends on $H_f$ and the
initial state, here taken as the groundstate of $H_i$.
For the special case of vanishing $V_f(x)$ the dynamics is just the free
wave packet evolution  of the ground state
$|\psi_0\rangle$   of $\hat H_i$.
The description of quenches in the framework of classical statistical
mechanics \cite{Huang}  is also presented. A  comparison 
with the quantum mechanical results is presented in sections III and IV.

A complete analytical description of the dynamics is possible when
the initial and final Hamiltonians are of the harmonic oscillator type.
We also present results for linear potentials
$V_f(x)=-Fx$ in section III.

 In section IV  we assume that  $ V_i(x)$ is a square well potential
and  $ V_f(x)$ vanishes.
 We present a simple introduction to the concept
of Wigner functions \cite{Wigner,Case}. This
allows the description of the broadening of quantum mechanical
wave packets similar to the classical case.

\section{Time evolution after the quench}

In the following we mainly focus on the probability density

\begin{eqnarray}
\label{probdensity}
  \rho_{qm}(x,t)&=&
  \langle \psi(t)|\delta(x-\hat x) 
                    |\psi(t)\rangle~, \nonumber \\
&=& \int  \langle \psi(t)|\delta(x-\hat x) |x'\rangle \langle x'
    |\psi(t)\rangle dx'~, \nonumber \\
&=&  |\langle x|\psi(t)\rangle|^2~.
\end{eqnarray}
to find the particle at position $x$ at time $t$,
where we have introduced the index $qm$ as we later compare quantum
quenches with quenches in the framework of classical
statistical mechanics.

Instead of directly working
with the delta function in Eq. (\ref{probdensity}) we use its integral
representation
\begin{equation}
\label{probden}
\rho_{qm}(x,t)  =\frac{1}{2\pi}\int_{-\infty}^\infty e^{-ikx}
\langle \psi(t)| e^{ik\hat x} |\psi(t)\rangle dk~.
\end{equation}
This rarely used expression for $\rho_{qm}(x,t)$
turns out to be very useful in the following.

The usual approach to obtain  $\rho_{qm}(x,t)$ for times after the quench
is to
calculate $ |\psi(t)\rangle= e^{-i\hat H_f t/\hbar} |\psi\rangle $, using the
eigenstates of $\hat H_f$,
where $|\psi\rangle$ is the initial state before the quench.
This is described for the free time evolution after the quench in section IV.

However, it is sometimes useful
to work in the Heisenberg picture in which operators acquire
the time dependence
$\hat A(t)=e^{i\hat H_f t/\hbar}\hat A  e^{-i\hat H_f t/\hbar}$.
Expanding $e^{ik\hat x}$ in Eq. (\ref{probden}) and using $\hat x^n(t)
=(\hat x(t) )^n$ one obtains
\begin{eqnarray}
\label{Heisenberg2}
\rho_{qm}(x,t) & =&\frac{1}{2\pi}\int_{-\infty}^\infty e^{-ikx}
                    \langle \psi| e^{ik\hat x(t)} |\psi\rangle dk~, \nonumber \\
  &=& \langle \psi|\delta (x-\hat x(t))|\psi \rangle~.
\end{eqnarray}

In classical statistical mechanics
for a single particle the initial state at time
$t=0$ is described by the
phase space probability density $\rho_0(x_0,p_0)$ for the initial
position $x_0$ and momentum $p_0$ of the particle \cite{Huang}.
The average classical particle density at finite times is given by
\begin{eqnarray}
 \label{classicalresult}
  \rho_{cl}(x,t) &=& \int \delta(x-x_{x_0,p_0}(t))\rho_0(x_0,p_0)dx_0dp_0
\nonumber \\
      &\equiv& \langle  \delta(x-x_{x_0,p_0}(t)) \rangle_{cl}   ~,          
\end{eqnarray}
where $ x_{x_0,p_0}(t)$ is the particle trajectory.
For the comparison with the quantum mechanical
results in section III we also make use of the
integral representation of the delta function
\begin{equation}
 \label{classical2}
 \rho_{cl}(x,t)=\frac{1}{2\pi}\int  e^{-ikx}
  \langle e^{ikx_{x_0,p_0}(t)} \rangle_{cl} dk~.
\end{equation}
If the particle is
in thermal equilibrium with a bath at temperature $T$
before the quench the
initial 
probability density $\rho_0(x_0,p_0)$   is proportional
to $e^{-\beta(p_0^2/2m+V_i(x_0))}$,
with $\beta=1/k_BT$.
This is an example for
a factorized initial distribution
$\rho_0(x_0,p_0)=\rho_0(x_0)\tilde\rho_0(p_0)$.

\section{Harmonic oscillators}

\subsection{The shifted harmonic potential}

The Hamiltonian of a one-dimensional harmonic oscillator reads
\begin{equation}
  \label{Hho}
  \hat H =\frac{1}{2m}\hat p^2+\frac{\lambda}{2}\hat x^2 ,
\end{equation}
with $\lambda$ the spring
constant. The angular frequency
of the oscilator is given by $\omega_0=\sqrt{\lambda/m}$.

The quantum mechanical description of the harmonic oscillator is especially
simple using the ladder operators. 
One defines 
the lowering operator $\hat a$ and its adjoint $\hat a^\dagger$
\begin{equation}
\label{defa} 
\hat a =\sqrt{\frac{m\omega_0}{2\hbar}}\hat x +
\frac{i}{\sqrt{2m\hbar \omega_0}}\hat p~;~~ \hat a^\dagger =
\sqrt{\frac{m\omega_0}{2\hbar}}\hat x -
\frac{i}{\sqrt{2m\hbar \omega_0}}\hat p~,
\end{equation}
which obey the commutation relation $[\hat a, \hat a^\dagger ]=\hat 1$.
 The position operator $\hat x$ and the 
 momentum operator $\hat p$ read in terms of $\hat a$ and $\hat a^\dagger$
 \begin{equation}
\label{xp}
\hat x =\sqrt{\frac{\hbar}{2m\omega_0}}\left (\hat a^\dagger+ \hat a
\right);~~\hat p=i\sqrt{\frac{m\hbar \omega_0}{2}}\left(\hat a^\dagger
 -\hat a \right ).
\end{equation}
 The ground state $|0\rangle$  is annihilated by $\hat a$, i.e.
 $\hat a|0\rangle=0$ holds.

 As $\hat H_i$ we use the Hamiltonian in
 Eq.(\ref{Hho}) with $\lambda \to \lambda_i$, and its  ground state
 $|0\rangle_i$ as the initial state.
 Averages of operators $\hat A$ in this ground state we denote by
 $\langle \hat A \rangle $. Then 
$\langle \hat x \rangle =0$
holds as well as $\langle \hat p\rangle=0$.

As the final potential we use
\begin{equation}
  \label{Vfinal}
  V_f(x)=-Fx+\frac{\lambda_f}{2}x^2=\frac{\lambda_f}{2}
  \left( x-\frac{F}{\lambda_f}\right)^2-\frac{F^2}{2\lambda_f}~,
\end{equation}
 i.e., a harmonic oscillator
with spring constant $\lambda_f$ shifted to
the position $a_F=F/\lambda_f$ and angular
frequency $\omega_f=\sqrt{\lambda_f/m}$.

The equations of motion and their solutions
for the operators $\hat x(t)$
  and $\hat p(t)$ are identical in form to the classical ones for $x(t)$ and
  $p(t)$.
  In the quantum mechanical case the classical initial conditions
  $x_0$ and $p_0$
  are replaced by $\hat x =\hat x(0)$ and $\hat p =\hat p(0)$.
  With the definition $\Delta \hat A \equiv
  \hat A-\langle \hat A \rangle$
  one obtains
 \begin{equation}
    \label{Deltax1}
    \langle \hat x(t)\rangle=a_F(1- \cos{(\omega_f t)}), ~~~
    \langle \hat p(t)\rangle =ma_F\omega_f \sin{\omega_f t} 
  \end{equation}
and
  \begin{eqnarray}
    \label{solutions}
    \Delta \hat x(t)&=&\hat x \cos{\omega_f t}+
                 \hat p \sin{(\omega_f t)}/m\omega_f \\    
    \Delta
    \hat p(t)&=&-\hat x m\omega_f \sin{
                \omega_f t}+\hat p \cos{\omega_f t}~.
  \end{eqnarray}
 
  In order to calculate $\rho_{qm}(x,t)$ it is useful to express
  $\Delta \hat x(t) $
  as a linear combination of the ladder operators
  corresponding to the
  harmonic oscillator $\hat H_i$
  \begin{equation}
    \label{Deltax}
 \Delta   \hat x(t)  =
    \alpha(t)\hat a^\dagger+\alpha(t)^*\hat a~,
  \end{equation}
  where using Eq. (\ref{xp}), $\alpha(t)$ is given by
  \begin{equation}
    \alpha(t)=\sqrt{\frac{\hbar}{2m\omega_i}}\left(\cos{\omega_f t}
    +i\frac{\omega_i}{\omega_f}\sin{\omega_f t}\right)~.
  \end{equation}
  Using $\hat a|0\rangle=0$ and $[\hat a,\hat a^\dagger]=\hat 1$
  the expectation value of $(\Delta \hat x(t))^2$
  in the ground state
  of $\hat H_i$ is given by $|\alpha(t)|^2$
  \begin{eqnarray}
    \label{widthx}
    \langle (\Delta \hat x(t))^2 \rangle  & =&|\alpha(t)|^2  \\
    &=&   \frac{\hbar}{2m\omega_i}\left( \cos^2{\omega_f t}+
        \left( \frac{\omega_i}{\omega_f}\right)^2\sin^2{\omega_f t}\right)
        \nonumber
  \end{eqnarray}
  In order to calculate  $\rho_{qm}(x,t)$ we use the
 Baker-Hausdorff-identity \cite {Merzb}. 
 It states that if $[\hat A,[\hat A,\hat B]]=0=[\hat B,[\hat A,\hat B]]$
 \begin{equation}
 \label{BH}
e^{\hat A +\hat B} =e^{\hat A}e^{\hat B}e^{-\frac{1}{2}[\hat A,\hat B]}~.
\end{equation}
For operators $\hat A$ and $\hat B$  linear in the ladder
operators the requirements are fulfilled.
Therefore the expectation value in Eq. (\ref{Heisenberg2})
can easily be calculated.  In the
integrand in  Eq. (\ref{Heisenberg2}) we write
\begin{equation}
e^{-ikx} e^{ik\hat x(t)}=  e^{-ik(x-\langle \hat x(t)\rangle)}e^{ik\Delta\hat x(t)}
\end{equation}
and evaluate the expectation value of $e^{ik\Delta\hat x(t)}$ using equations
 (\ref{Deltax}) and (\ref{BH})
\begin{eqnarray}
  \label{roqm}
  \langle e^{ik\hat\Delta x(t)}\rangle &=& \langle e^{ik\alpha(t)\hat
         a^\dagger+ik \alpha(t)^*\hat a}\rangle \nonumber \\
 &=& \langle e^{ik\alpha(t)\hat a^\dagger }e^{ik\alpha(t)^* \hat a }\rangle
e^{-k^2 |\alpha(t)|^2/2 }\nonumber \\
&=&  e^{-k^2 |\alpha(t)|^2/2}   ~.
\label{eikx}
\end{eqnarray}
Because of $\hat a|0\rangle_i=0$ which implies
  $_i\langle 0|\hat a^\dagger=0 $ the expectation value in the second equality
equals $1$.
Putting this into the upper part of  Eq. (\ref{Heisenberg2})
the Gaussian integration can be
performed and one obtains
\begin{equation}
  \label{xresult}
 \rho_{qm}(x,t)=
  \frac{1}{\sqrt{2\pi\langle (\Delta\hat x(t))^2\rangle}}
  \exp{\left(  -\frac{(x- \langle \hat x(t)\rangle )^2}
    {2 \langle (\Delta\hat x(t))^2\rangle}\right)}~.
\end{equation}
An analogous result is obtained for the momentum probability
distribution 
\begin{equation}
  \langle \delta(p-\hat p(t)\rangle=
  \frac{1}{\sqrt{2\pi\langle (\Delta\hat p(t))^2\rangle}}
  \exp{\left(-\frac{(p- \langle \hat p(t)\rangle )^2}
    {2 \langle (\Delta\hat p(t))^2\rangle}\right)}~.
\end{equation}
with the width determined by
\begin{equation}
  \langle (\Delta \hat p(t))^2\rangle=\frac{m\hbar \omega_i}{2}
  \left( \cos^2{\omega_f t}+
 \left(\frac{\omega_f}{\omega_i}\right)^2\sin^2{\omega_f t}\right)~.
\end{equation}

Fig. 1 shows the time dependence of $\langle \hat x(t) \rangle$,
$\Delta x(t)=\sqrt{  \langle (\Delta \hat x(t))^2\rangle}$, 
$\Delta p(t)=\sqrt{  \langle (\Delta \hat p(t))^2\rangle}$ and the
uncertainty product.
The initial wave packet makes an oscillatory motion around the center
of the harmonic potential in $H_f$.
The width $\Delta x(t)$ and $\Delta p(t)$ oscillate with double frequency
and the uncertainty product  $\Delta x(t)\Delta p(t)$ equals $\hbar/2$
at positions $0,a_F$ and $2a_F$ and is larger at the intermediate positions.

\begin{figure}
\label{squeez}
\centering
\includegraphics[width=6cm,height=5cm]{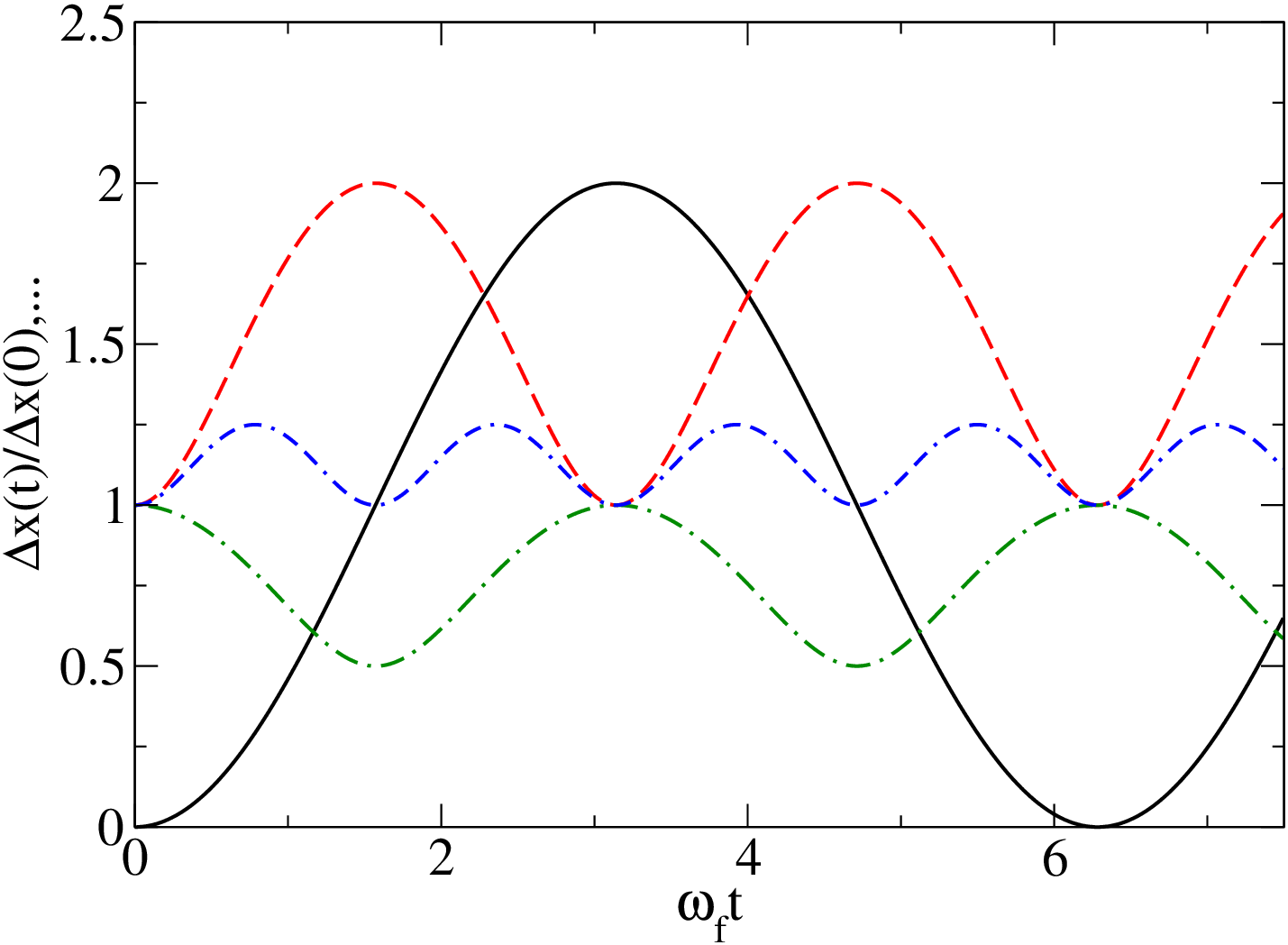}
\vspace{0.5cm}
\caption {Time dependence of  $\langle \hat x(t) \rangle/a_F$ (solid line),
  $ \Delta x(t)/\Delta x(0)$  (dashed line), $ \Delta p(t)/\Delta p(0)$
  ($- \cdot -\cdot -$), and 
  the uncertainty product $2\Delta x(t)\Delta p(t)/\hbar$
  ($ - - \cdot - -  \cdot$) as a function of
  $\omega_f t$  for $\omega_f/\omega_i=1/2$.}
\end{figure}
The behaviour of $ \rho_{qm}(x,t) $ in Eq. (\ref{xresult})
is well known. 
 It is usually obtained by calculating $\langle x|0(t)\rangle_i$
introducing the concepts of coherent and squeezed states \cite{Merzb,Ball}
as a typical quantum mechanical behaviour. The solution presented here 
without introducing these states is simpler.

Surprisingly  Eq. (\ref{xresult})
can also be obtained purely classically.
To show this we consider the classical quench dynamics
switching from $H_i$ to $H_f$
when $\rho_{cl}(x_0,p_0)=\rho_0(x_0)\tilde\rho_0(p_0)$ and $\rho_0(x_0)$
and $\tilde\rho_0(p_0) $ are both Gaussians. Then the integration in Eq. (\ref
{classical2}) can be performed analytically
by first calculating
\begin{equation}
  \langle e^{ikx_{x_0,p_0}(t)} \rangle_{cl}=e^{ik\langle x(t) \rangle }
    \langle  e^{ik(x_0\cos{\omega_ft}+p_0\sin{(\omega_f t)}/m\omega_f)}\rangle_{cl}
 \end{equation}   
 by Gaussian integrations. The remaining $k-$integration is also
 Gaussian and one obtains
\begin{equation}
 \rho_{cl}(x,t)=
  \frac{1}{\sqrt{2\pi\langle (\Delta x(t))^2\rangle}_{cl}}
  \exp{\left(  -\frac{(x- \langle  x(t)\rangle )^2}
    {2 \langle (\Delta x(t))^2\rangle_{cl}}\right)}
\end{equation}
with
\begin{equation}
  \langle (\Delta x(t))^2\rangle_{cl}=\langle x_0^2\rangle \cos^2({\omega_ft})
  +\langle p_0^2\rangle \sin^2({\omega_ft})/(m\omega_f)^2~.
\end{equation}
For the case of the canonical ensemble discussed after
Eq. (\ref{classical2}) one has
$\langle p_0^2 \rangle/\langle x_0^2 \rangle =
m^2\omega_i^2$ and the time dependence of $\langle (\Delta x(t))^2\rangle_{cl}$,
apart from the prefactor
is the same as in Eq.(\ref{widthx}).
For the special temperature choice
$k_BT=\hbar \omega_i/2$,
also the prefactor is the same.  This corresponds to the initial condition
\begin{equation}
  \label{clqm}
  \rho_0(x_0,p_0)= \psi_0(x_0)^2\tilde \psi_0(p_0)^2~,
\end{equation}
where $\psi_0(x_0)=\langle x_0|\psi_0 \rangle$ is the real wave function 
for the Gaussian initial state $|\psi_0 \rangle =|0\rangle_i$
considered in this section
and  $\tilde \psi_0(p_0)=\langle p_0|\psi_0 \rangle$  the
corresponding real
Gaussian momentum amplitude.\\

It is left as an exercise to repeat the caclulation of $\rho_{qm}(x,t)$
for $|1\rangle_i= \hat a^\dagger |0\rangle_i$
and $|2\rangle_i=(\hat a^\dagger)^2|0\rangle_i /\sqrt 2$ as the initial state
by expanding
the exponential functions in the expectation value in the second equality in
Eq. (\ref{roqm}). The results for the special case $\hat H_f=\hat T$
can be found in reference 7. There the free time evolution after the
quench for the initial state $c\hat p^2 |0\rangle_i$ is also discussed.
It is a special linear combination of   $|0\rangle_i$ and $|2\rangle_i$.
Results for the general linear combination of these two states
are presented in reference 8.

\subsection{The linear potential}

We next discuss the case $\lambda_f=0$ and a finite value of $F$, i.e.
the linear potential $V_f(x)=-Fx$. 
Performing the limit $\omega_f \to 0$ in  
Eq.(\ref{Deltax1}) leads to
\begin{equation}
  \label{linear1}
\langle (\Delta \hat x(t))^2\rangle = \langle \hat x^2\rangle (1+\omega_i^2t^2)
\end{equation}
independent of the value of $F$ which only appears in the expression
for $\langle \hat x(t) \rangle$ in Eq. (\ref{Deltax1}), given by
\begin{equation}
  \label{linear2}
  \langle \hat x(t) \rangle_F=F t^2/2m~.
\end{equation}
The result for $|\langle x|\psi(t)\rangle |^2$ for the
linear potential is given by Eq. (\ref{xresult}) with the results of
Eq. (\ref{linear1}) and
Eq. ({\ref{linear2}) inserted. The time dependence of the broadening
  is identical to the free particle case.

  This holds for linear potentials for arbitrary initial states
  $|\psi(0)\rangle$. The solution of the Heisenberg
  equation of motion is given by
\begin{equation}
  \label{wep1}
  \hat x(t)=\hat x+\hat pt/m +F t^2/2m~.
\end{equation}
Putting this into the second line of Eq. (\ref{Heisenberg2}) yields
\begin{equation}
  \label{wep2}
  |\langle x|\psi(t)\rangle_F|^2=
  |\langle x-\langle \hat x(t) \rangle_F |\psi(t)\rangle_{F=0}|^2
\end{equation}
This is in accord with the (weak) equivalence
principle \cite{Holstein,Nauenberg},
which states that all laws of a freely falling particle are the same as in
an unaccelarated reference frame.
The presented proof for the broadening of wave packets using the
Heisenberg picture by directly adressing the measurable probability density
$|\langle x|\psi(t)\rangle|^2$
is much simpler than using the Schr\"odinger picture
and calculating
$\langle x|\psi(t)\rangle$ first.
\cite{Holstein,Nauenberg}.

In quantum mechanics textbooks the factor multiplying $t^2$
for the free case $F=0$ in Eq. (\ref{linear1}) is usually
expressed differently as the broadening of the Gaussian wave packet
is treated before the harmonic oscillator. Instead of $\omega_i^2$
the factor is written $(\hbar/(2m\langle \hat x^2\rangle))^2 $

For the case
$\lambda_f=0$ and $F=0$ the free broadening of a Gaussian wave packet
using operator manipulations was 
presented in this journal recently \cite{Jim}.\\

\section {Free time evolution after the quench}

In the previous section we used the Heisenberg picture to calculate
the time dependence of the probability density $\rho_{qm}$ after the
quench.
  Here we 
use the Schr\"odinger picture to obtain $\rho_{qm}(x,t) $
for the quench in which  $V_f(x)\equiv
0$.

We begin with the  usual approach to obtain  $\rho_{qm}(x,t)$   
by calculating $ \langle x|\psi(t)\rangle$ and taking its absolute
square. In the second part of this section we discuss the additional
insight one can obtain by using Eq. (\ref{probden}) instead.

In the first approach one uses the eigenstates  of $\hat H_f$.
For the case $\hat H_f=\hat T$ those are given by
the momentum states $|p \rangle$.
Inserting the unit operator expressed
in terms of the momentum states and  using $\langle x|p\rangle=
 e^{ipx/\hbar}/\sqrt{2\pi\hbar}  $ yields

\begin{equation}
  \label{xwavepacket}
  \langle x|\psi(t)\rangle=\frac{1}{\sqrt{2\pi\hbar}} \int_{-\infty}^\infty
   e^{ipx/\hbar}  \langle p|\psi \rangle e^{-i\epsilon_pt/\hbar}dp~,
\end{equation}
with $\epsilon_p=p^2/2m$ and the momentum representation
\begin{equation}
  \label{pwavepacket}
\tilde \psi(p)\equiv \langle p|\psi \rangle
  =\frac{1}{\sqrt{2\pi\hbar}} \int_{-\infty}^\infty
   e^{-ipx/\hbar} \langle x|\psi \rangle dx~,
\end{equation}
 of the arbitrary initial state $|\psi\rangle$.

 If the wave function
$\psi (x)=\langle x|\psi \rangle$ is Gaussian the same holds for
$\tilde \psi(p)$ and the integration
in Eq. (\ref{xwavepacket}) can be
performed analytically.
This is presented in almost all quantum mechanics textbooks. For generic
$\psi(x)$ the integration in Eq. (\ref{xwavepacket}) has to
be performed numerically.

In this section we take as the initial state $|\psi_0 \rangle $
 the groundstate
of an attractive square well potential $V_i(x)=V_i\Theta(a-|x|)$,
 with $\Theta$ the step function and $V_i<0$.
Its wave function is given by
\begin{equation} \label{squarewell}
\psi_0(x)=\begin{cases}
c_0\cos(k_0x)/\cos(k_0a) & |x| \leq a \\
c_0e^{-\kappa (|x|-a)} & |x| \geq a
\end{cases}
\end{equation}
with $k_0=\sqrt{2m(E_0-V_i)}/\hbar$ and the groundstate energy $E_0$   .

In order to have continuous first derivatives at $x=\pm a$
\begin{equation}
  \kappa =k_0\tan(k_0a)
\end{equation}
has to hold with  $\tan(k_0a)>0$. The
normalization constant is given by
$c_0=[1/\kappa +a(1+\kappa^2/k_0^2)+\kappa/k_0^2]^{-1/2}$. The dependence
of $E_0$ on $V_i$ is irrelevant in the following.

There are two interesting limiting cases:

i) The limit $a\to 0, V_i \to -\infty $ with arbitrary $\kappa >0$,
i.e. an attractive delta potential
with the ground state wave function
$ \psi_0(x)=\sqrt \kappa e^{-\kappa|x|}$.

ii) The infinitely deep potential: For
$ak_0 \to \pi/2$
  the ratio $\kappa/k_0$
tends to infinity and one obtains
$ \psi_0(x)=\cos(k_0x)\Theta(a-|x|)/\sqrt a$.

The momentum representation of $|\psi_0\rangle$
can be calculated analytically.
Using Eqs. (\ref{pwavepacket}) and (\ref{squarewell})
one obtains with $p=\hbar k$
\begin{equation}
  \label{psivonk}
  \tilde\psi_0(\hbar k)=\frac{2c_0}{\sqrt{2\pi \hbar}}(\kappa \cos ka-k\sin ka)
  \left( \frac{1}{k^2+\kappa^2} -\frac{1}{k^2-k_0^2}\right)~.
\end{equation}
Inserting this result into the integral in Eq.(\ref{xwavepacket}) it can be
calculated numerically to obtain
$\langle x|\psi(t)\rangle$. Its absolute value squared is shown in Fig. 2 for
$k_0a=\pi/2$, i.e. the ground state of the
infinitely deep well, for four different times.
It shows that the probability density
to find the particle at the origin is larger
for $t=0.14 t_0$ than at the initial time $t=0$, where  $t_0=ma^2/\hbar$.
For $t=0.07 t_0$
it has a minimum at the origin.

\begin{figure}
\label{piheps}
\centering
\includegraphics[width=6cm,height=5cm]{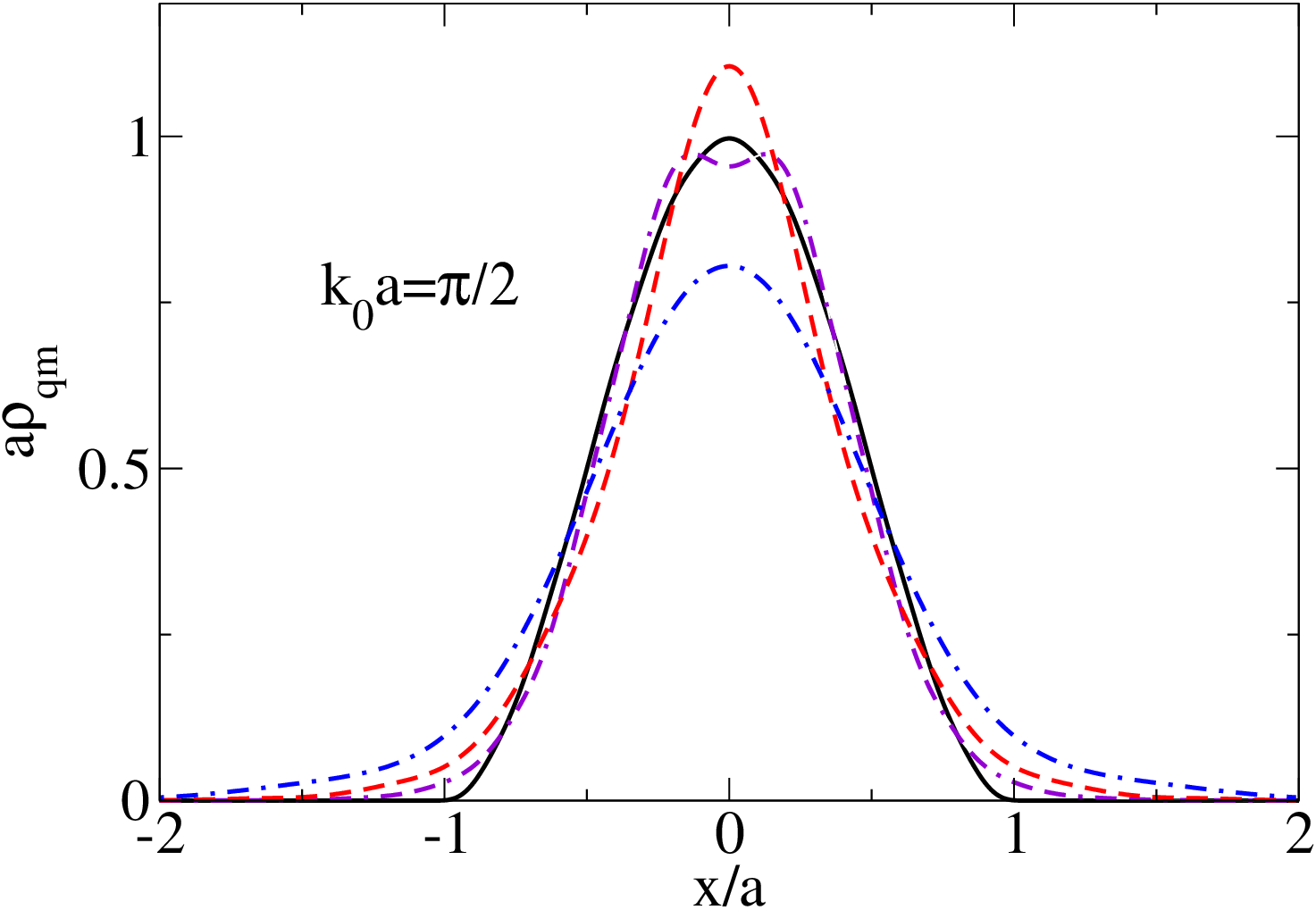}
\vspace{0.5cm}
\caption {Probability density $\rho_{qm}$ times $a$ as a function of $x/a$
  for four different times: $t=0$ (solid line), $t=0.07t_0$ 
  ($ - - \cdot - -  \cdot$)
  $t=0.14t_0  $ (dashed line)
  and   $t=0.28t_0  $  ($- \cdot -\cdot -$).   Note that $a\rho_{qm}(0,0.14)$
  is larger than  $a\rho_{qm}(0,0)$.}
\end{figure}

To elucidate this surprising
effect in more detail $a\rho_{qm}$ is shown for $x=0$
as function of $t/t_0$ for three
different values of $k_0a$ in Fig. 3.
\begin{figure}
  \label{kurzzeit}
\centering
\includegraphics[width=6cm,height=5cm]{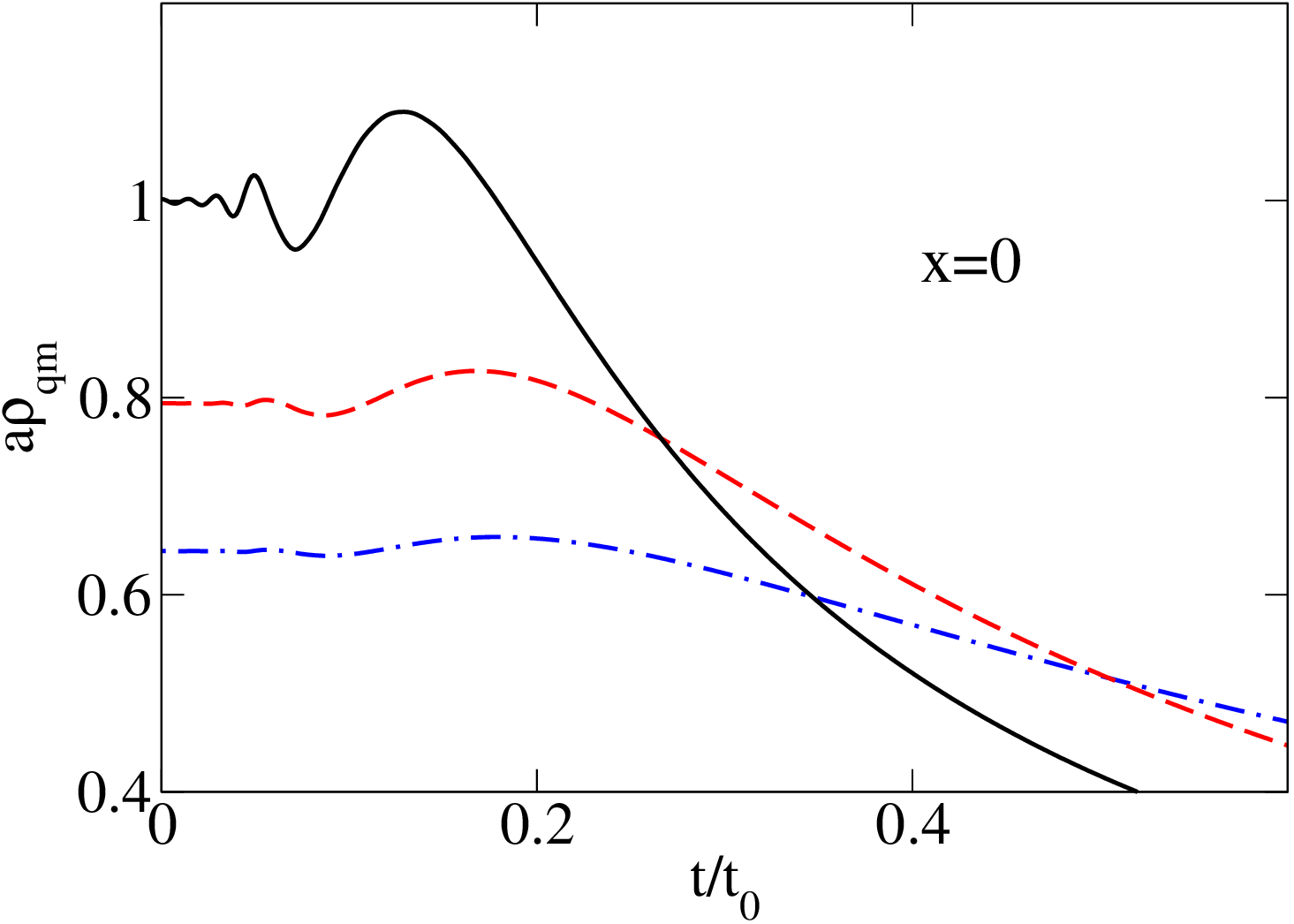}
\vspace{0.5cm}
\caption {Probability density  $\rho_{qm}$ times $a$
  for $x=0$ as a function of $t/t_0$
  for three different values of $k_0a$:
  $k_0a=\pi/2$, which corresponds to the
  infinitely deep well (solid line), $k_0a=\pi/2.5$ (dashed line)
  and $k_0a=\pi/3 $ ($- \cdot -\cdot -$) .}
\end{figure}
The short time oscillatory behaviour is more pronounced
when the well is deeper.

The fact that the probability to find the particle at the origin
for a $t>0$ is larger than in the initial state is a purely
quantum mechanical effect. It is easy to see that this cannot happen
in the classical case when the probablity to find the particle at
$t=0$ has its maximum at $x_0=0$.
For a free particle the trajectory is given by $x_{x_0p_0}(t)=x_0+p_0t/m$ and
using Eq. (\ref{classicalresult})
one obtains
\begin{equation}  
\label{classical}
\rho_{cl}(x,t)=\int \rho_0(x-\frac{p_0t}{m},p_0)dp_0 ~.
\end{equation}
In order to be as close as possible to the quantum mechanical case for
a general initial state $|\psi \rangle$  the phase
space density $\rho_0(x_0,p_0)$ should yield
the quantum mechanical space probability $|\langle x_0|\psi\rangle|^2$
by integration over $p_0$
and the momentum probability $|\langle p_0|\psi\rangle|^2$
by integration over $x_0$.
 By chosing
 \begin{equation}
   \label{clinitial}
   \rho_0(x_0,p_0)= |\langle x_0|\psi\rangle|^2|\langle p_0|\psi\rangle|^2~.
\end{equation}
this is obviously fulfilled. We call this the
 ``classical approximation'' in quotation marks as the dynamics
 using Eq. (\ref{classical}) is classical, but this initial
condition involves quantum mechanical probability densities.

For $|\psi \rangle$ the ground state
of the square well potential $|\langle x_0|\psi\rangle|^2$ has its maximum
at $x_0=0$.
 For $t>0$ one has $ \rho_0(-\frac{p_0t}{m},p_0)< \rho_0(0,p_0)$ which 
 using Eq. (\ref{classical}) implies
 $\rho_{cl}(0,t) < \rho_{cl}(0,0)$.

 It turns out that the ``classical
 approximation'' works rather well for larger times. This is shown
 in Fig. 4 for $k_0a=\pi/2$.

 For a better understanding of this we show that the ``classical
 approximation'' gives the exact result for the time dependence 
 of the quantum mechanical width of the wave packet.  With the initial
 condition in Eq.(\ref{clinitial}) one obtains $\langle (x(0))^n\rangle_{cl}
 =\langle (\hat x(0))^n\rangle_{qm}$ and $\langle (p(0))^n\rangle_{cl}
 =\langle (\hat p(0))^n\rangle_{qm}$. As we consider initial states
 $|\psi \rangle $ with even wave functions $\langle x(0)\rangle_{cl}=0$
  and $\langle (x(0) p(0)\rangle_{cl}=0$ holds.
  This leads to
 \begin{equation}
   \label{clwidth}
   \langle (\Delta x(t))^2\rangle_{cl}  =
   \langle \hat x^2 \rangle_{qm} +t^2\frac{\langle \hat p^2\rangle_{qm}}{m^2} ~.
 \end{equation}

 To calculate the
 expectation value
 $\langle \hat x \hat p \rangle +\langle \hat p \hat x\rangle$
 for a square integrable wave function $\psi(x)$ one can use that the integral
 over the derivative of $x|\psi(x)|^2$ vanishes as
 $x|\psi(x)|^2\to 0$ for $x \to \pm \infty$.
 For $\psi(x)=f(x)e^{i\phi(x)}$ with $f$
 and $\phi$ real functions one obtains (exercise)
 \begin{equation}
   \langle \hat x \hat p \rangle +\langle \hat p \hat x\rangle
   =2\hbar \int_{-\infty}^\infty x\phi'(x)f^2(x)dx~.
 \end{equation}
 Therefore the sum vanishes for real $\psi(x)$.
  This leads to
 \begin{equation}
   \label{clwidth}
   \langle (\Delta \hat x(t))^2\rangle_{qm}  =
   \langle (\Delta x(t))^2\rangle_{cl} ~.
 \end{equation}

In order to discuss the long time behaviour of   $\rho_{cl}(x,t)$ one
 can alternatively perform the $p_0$ integration
in Eq. (\ref{classicalresult}) first or change the integration variable in
Eq. ({\ref{classical}), both leading to 
\begin{equation}
\label{cllongtime}
\rho_{cl}(x,t)=\frac{m}{t}\int \rho_0(x_0,\frac{mx}{t}-
\frac{mx_0}{t})dx_0 ~.
\end{equation}
If  $|\psi(x_0)|^2$ in Eq. (\ref{clinitial}) decays fast enough
like for the square well potential Eq. (\ref{squarewell})
the term $mx_0/t$ in the integrand can be neglected in the
long time limit leading to
\begin{equation}
  \label{cllt}
  t\rho_{cl}(x,t) \to m |\tilde \psi(\frac{mx}{t})|^2
\end{equation}
for  finite $x/t$.\\

\begin{figure}
\label{clversusqm}
\centering
\includegraphics[width=6cm,height=5cm]{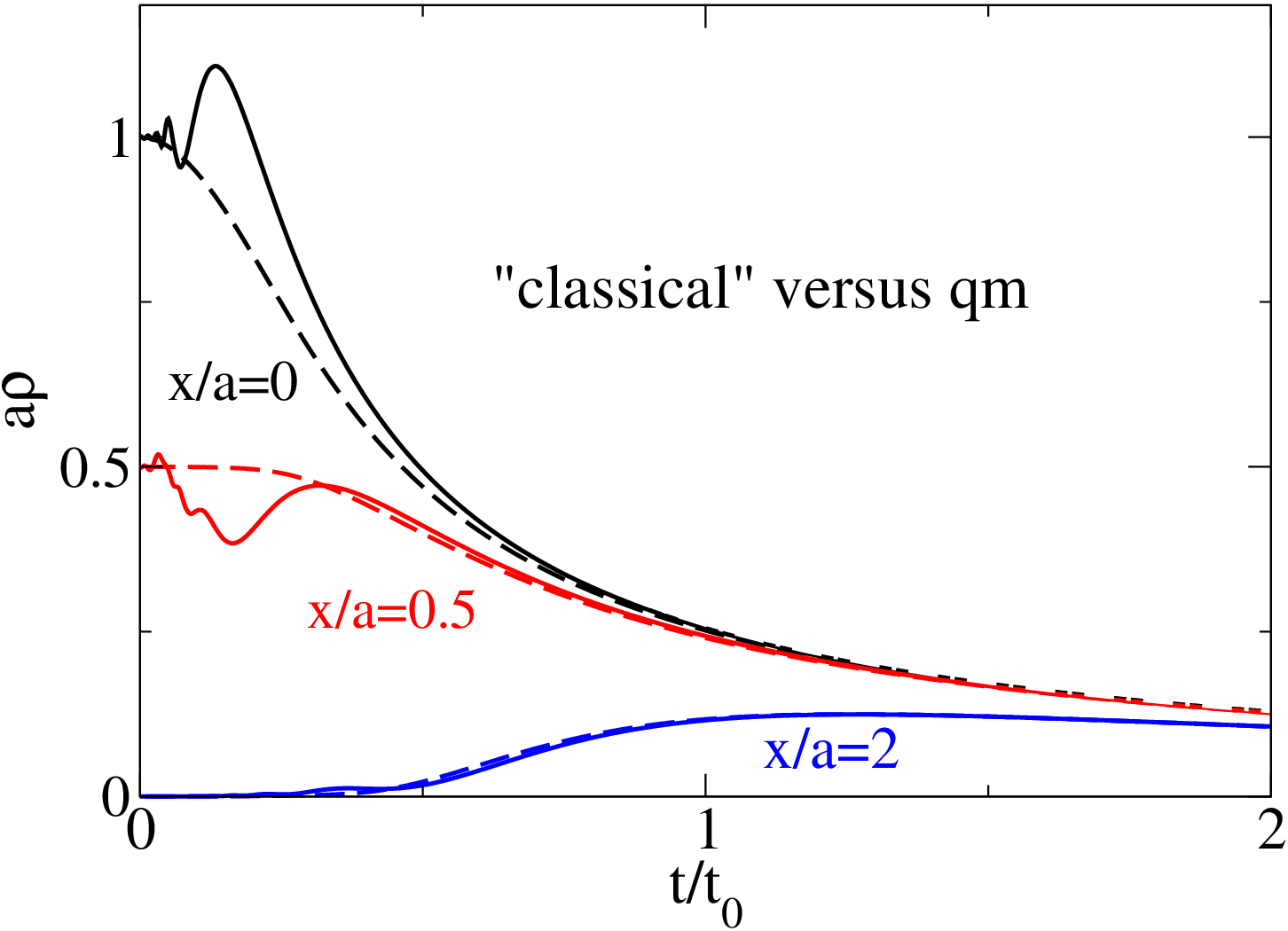}
\vspace{0.5cm}
\caption {Probability density $\rho_{qm}$ times $a$ for $k_0a=\pi/2$
  as a function of $t/t_0$
  for three different values of $x/a$ (solid lines), compared to the
  ``classical approximation'' described in the text (dashed lines).}
\end{figure}

\begin{figure}
\centering
\includegraphics[width=6cm,height=5cm]{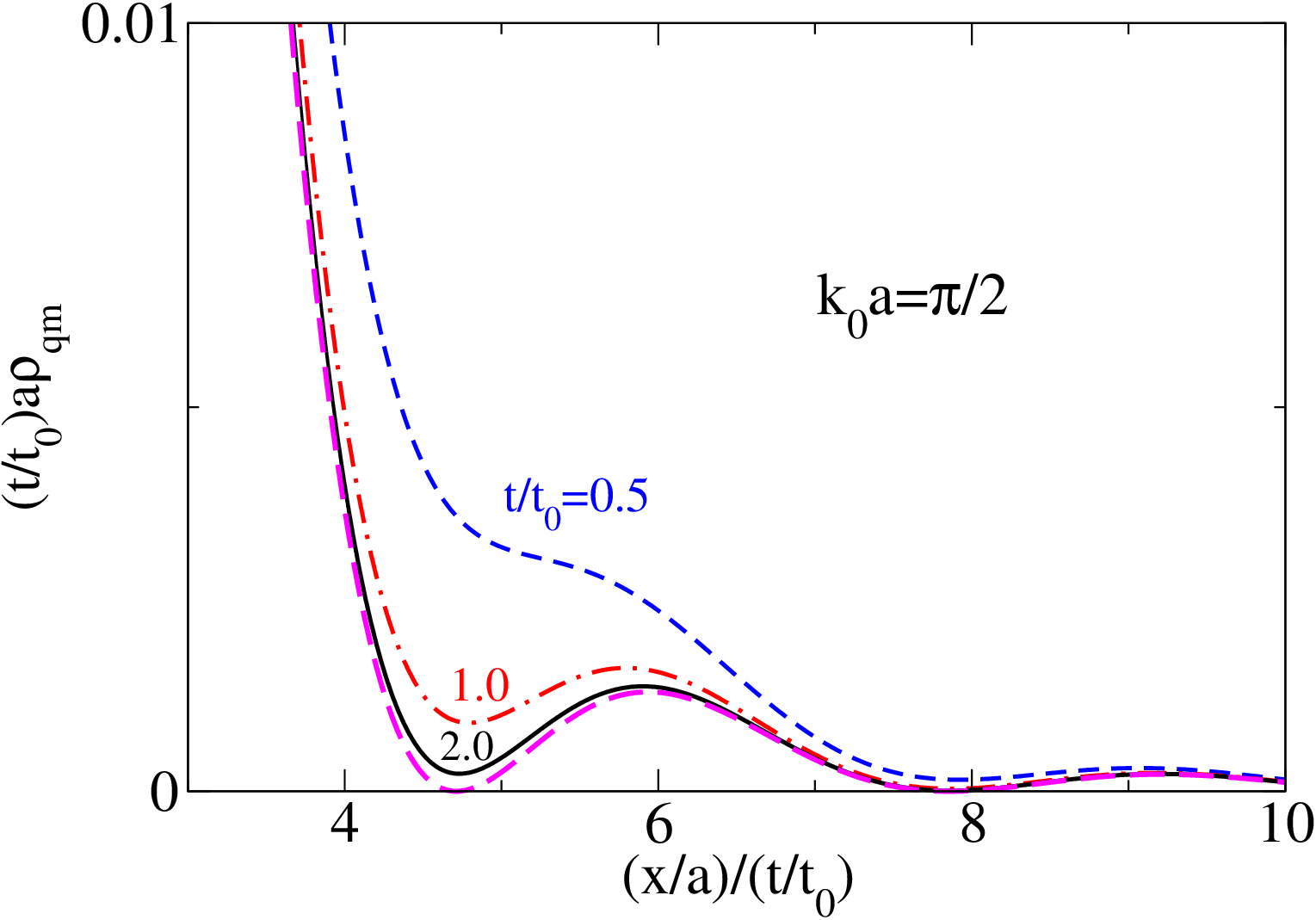}
\vspace{0.5cm}
\caption {Long time behaviour of $(t/t_0)a\rho_{qm}$ for the infinitely deep
  square well as a function of $(x/a)/(t/t_0)$ for different values of
  $t/t_0$.
  It quickly approaches $\hbar|\tilde
 \psi((x/a)/(t/t_0) )|^2/a$ (long dashed curve).}
\end{figure}

In Fig. 5 we show numerical results for the
quantum mechanical long time behaviour using
Eqs. (\ref{xwavepacket}) and (\ref{psivonk}).
The function $(t/t_0)a\rho_{qm}$ is shown  for various
values of $t/t_0$ as a function of $(x/a)/(t/t_0)$. It quickly approaches
$\hbar|\tilde \psi((x/a)/(t/t_0))|^2/a$, in agreement with Eq. (\ref{cllt}).\\

Before we present an explanation why
the quantum mechanical and the ``classical'' long time behaviour agree
we switch to  Eq. (\ref{probden})   to calculate $\rho_{qm}(x,t)$
without computing $\langle x|\psi(t)\rangle $ first.

For the later treatment of the free particle time dependence
it turns out to be useful to factorize $e^{ik\hat x}=e^{ik\hat x/2}
e^{ik\hat x/2}  $ and insert the unit operator expressed in
terms of the momentum
states in between.
Using $e^{ik\hat x/2}|p\rangle=|p+\hbar k/2\rangle $
as can be proven by multiplying both sides by $\langle x|$ this leads to 
\begin{equation}
e^{ik\hat x}=\int |p+\hbar k/2\rangle \langle p-\hbar k/2 | dp~.
\end{equation}
Putting this into Eq. (\ref{probden}) and changing the integration
variable $k=\tilde p/\hbar$ one obtains
\begin{equation}
  \label{Wintp}
\rho_{qm}(x,t)=   \int W(x,p,t)dp~,
\end{equation} 
with

\begin{equation}
  \label{Wignerp}
 W(x,p,t)=\frac{1}{2\pi\hbar}\int e^{-i\tilde px/\hbar}
  \langle \psi(t)|p+\tilde p/2\rangle
  \langle p-\tilde p/2 |\psi(t)\rangle d\tilde p~.
\end{equation}
This is a real function
as seen by changing the integration variable $\tilde p \to -\tilde p$.
Integration over $x$ gives    a factor $ 2\pi\delta(\tilde p/\hbar)$
leading to
\begin{equation}
  \int W(x,p,t)dx=|\langle  p|
  \psi(t)\rangle|^2~.
\end{equation}
  It is left as an exercise to show that after inserting unit operators in
  terms of
  position states in the integral in Eq. (\ref{Wignerp}) the function
  $W(x,p,t)$ can also be written as
  \begin{equation}
    \label{Wigx}
  W(x,p,t)=\frac{1}{2\pi\hbar}\int e^{-ipy/\hbar}\langle \psi(t)|x-y/2 \rangle
  \langle x+y/2 |\psi(t)\rangle dy~.
\end{equation}

\noindent In 1932 Wigner \cite{Wigner}
introduced the  quantum phase-space distribution $W(x,p,t)$
for arbitrary time dependent states $|\psi(t)\rangle$,
which has similarities
to the classical phase-space distribution \cite{Huang},
but allows to obain exact quantum mechanical results,
as shown above.  The higher dimensional generalization of the
Wigner function is widely
used as a tool in various areas, e.g. quantum optics \cite{Schleich,WFST}.
However, it is introduced only in a few quantum mechanics textbooks, e.g.
\cite{Merzb,Ball}.
A ``pedestrian'' introduction has been published in this journal \cite{Case}.

For  $W(x,p,0)$ to be a
probability density, i.e. it is positive
everywhere,  $\psi(x)$ has to be the exponential of a quadratic
polynomial \cite{Hudson}. 
A well known example is  a real-valued Gaussian.
The integration in Eq. (\ref{Wigx}) can then be easily performed and
one obtains $W(x,p,0)=\psi(x)^2\tilde \psi(p)^2$.

  The form of the Wigner function in Eq. (\ref{Wignerp}) is
  well suited to determine its time
dependence for   free particles  discussed in this section.
 The momentum states are the eigenstates for $\hat H_f=\hat T$.
 With $\epsilon_{p+\tilde p/2}-\epsilon_{p-\tilde p/2}=\tilde pp/m$
one obtains
\begin{equation}
  W(x,p,t)=W(x-\frac{pt}{m},p,0)~.
  \end{equation}
  If this is inserted into Eq. (\ref{Wintp}) it leads to
  \begin{equation}
    \label{Wigt}
 \rho_{qm}(x,t)=\int W(x-\frac{pt}{m},p,0)dp~.
  \end{equation}
  This result for $\rho_{qm}(x,t)$
  has the form as in the classical case in Eq. (\ref{classical}),
  with $W(x,p,0)$ instead of $\rho_0(x,p)$. Therefore the arguments leading
  to Eq. (\ref{cllt}) can be generalized to obtain
\begin{equation}
  \label{qmlt}
  t\rho_{qm}(x,t) \to  m| \tilde \psi(\frac{mx}{t})|^2~.
\end{equation}
if $W(x,p,0)$ has a well-localized $x$-dependence, which
is the case for the groundstate
of a deep square well potential.

The long time behaviour of the
broadening of wave packets has been discussed previously in two
papers in this journal \cite{Andrews,Mita}. In reference 7
the result  Eq. (\ref{qmlt}) is presented using the
free particle propagator, shortly discussed in the appendix.
The statement in reference 17,
that all wave packets become ``approximately
Gaussian'' after a long enough time, is critically discussed.

The Wigner function
for the ground state of the infinitely deep square well
can be obtained by simple integration using Eq. (\ref {Wigx})\cite{Thomas}
\begin{eqnarray}
  W(x,\hbar k,0)&=&\frac{ \Theta(a-|x|)
                            }{2\pi\hbar}
                       \Biggl[ \cos(\pi x/a)\frac{\sin(2k(a-|x|))}{ak}
  \nonumber \\
          &+&  \frac{ \sin((2ak+\pi)(1-|x|/a ))}{2ak+\pi}
                            \nonumber \\
    &+&  \frac{ \sin((2ak-\pi)(1-|x|/a))}{2ak-\pi}\Biggr]~.
\end{eqnarray}
It is shown in Fig. 6 as function 
of momentum for two different values of $x$. One can see that $W$ is negative
in some intervals. This is different for $  |\psi (0)|^2|\tilde \psi(p)|^2 $
shown as the dashed curve.

\begin{figure}
\label{Wignerx}
\centering
\includegraphics[width=6cm,height=5cm]{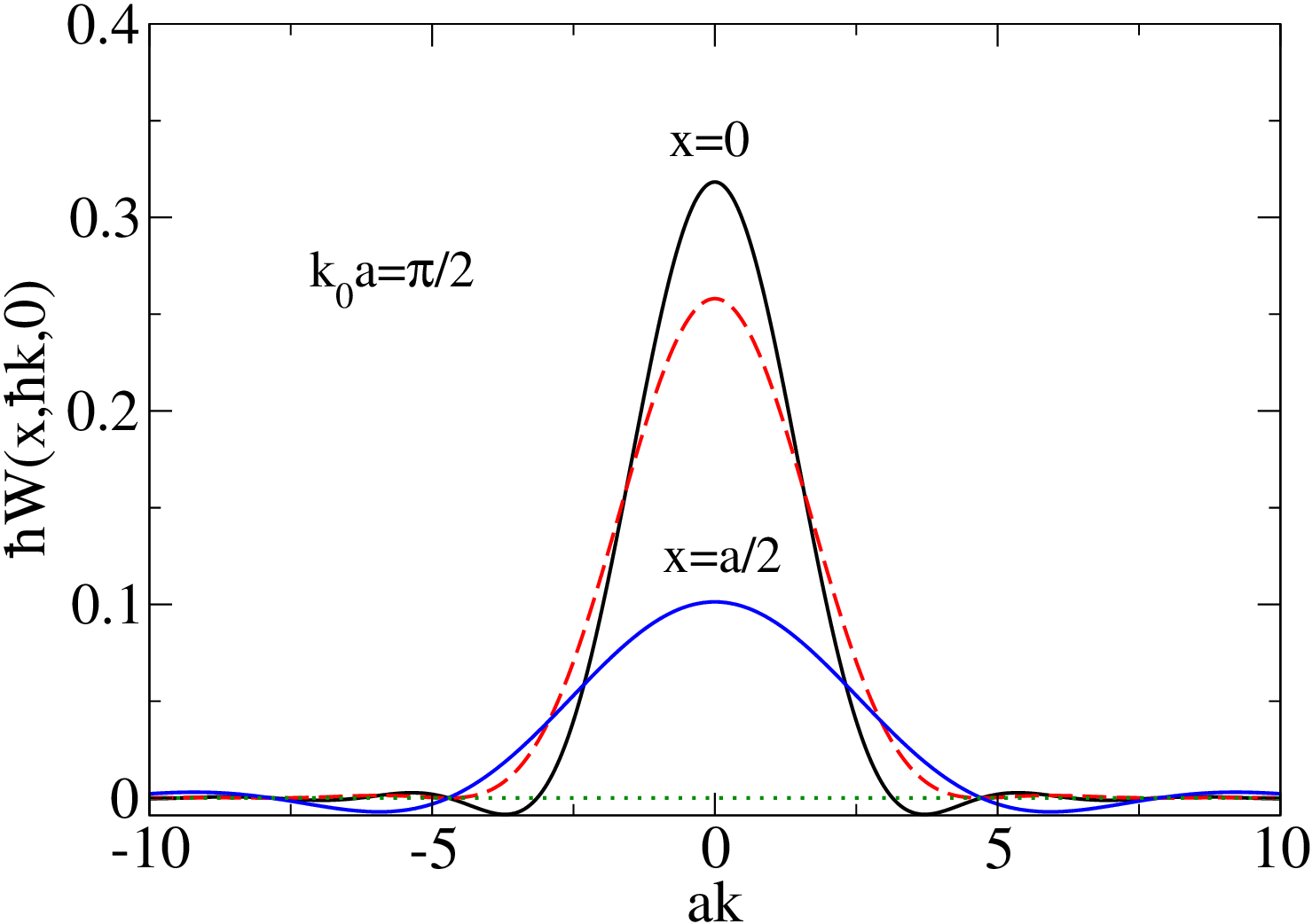}
\vspace{0.5cm}
\caption { Wigner function $ \hbar W(x,\hbar k,0)$ 
  as a funcion of $ak$ for $x=0$  and $x=a/2$.
  The dashed curve is the positive function
 $ \hbar|\psi (0)|^2|\tilde \psi(\hbar k)|^2$.   }
\end{figure}
The oscillatory behaviour of $\rho_{qm}(0,t)$ shown in Fig. 3 can be understood
without performing a numerical integration by plotting the integrand in
Eq. (\ref{Wigt})
$W(-\hbar kt/m,\hbar k,0)
=W(ka^2t/t_0,\hbar k,0)$
 (symmetric in $k$).  In Fig. 7 this is shown
for the times of the deepest minimum ($t/t_0\approx 0.071$) and highest
maximum ($t/t_0\approx 0.128$) of the full line in Fig 3.
 One clearly sees that the integral over the
  full curve yields a result larger than that of the dotted curve.
\begin{figure}
\label{Wignerx}
\centering
\includegraphics[width=6cm,height=5cm]{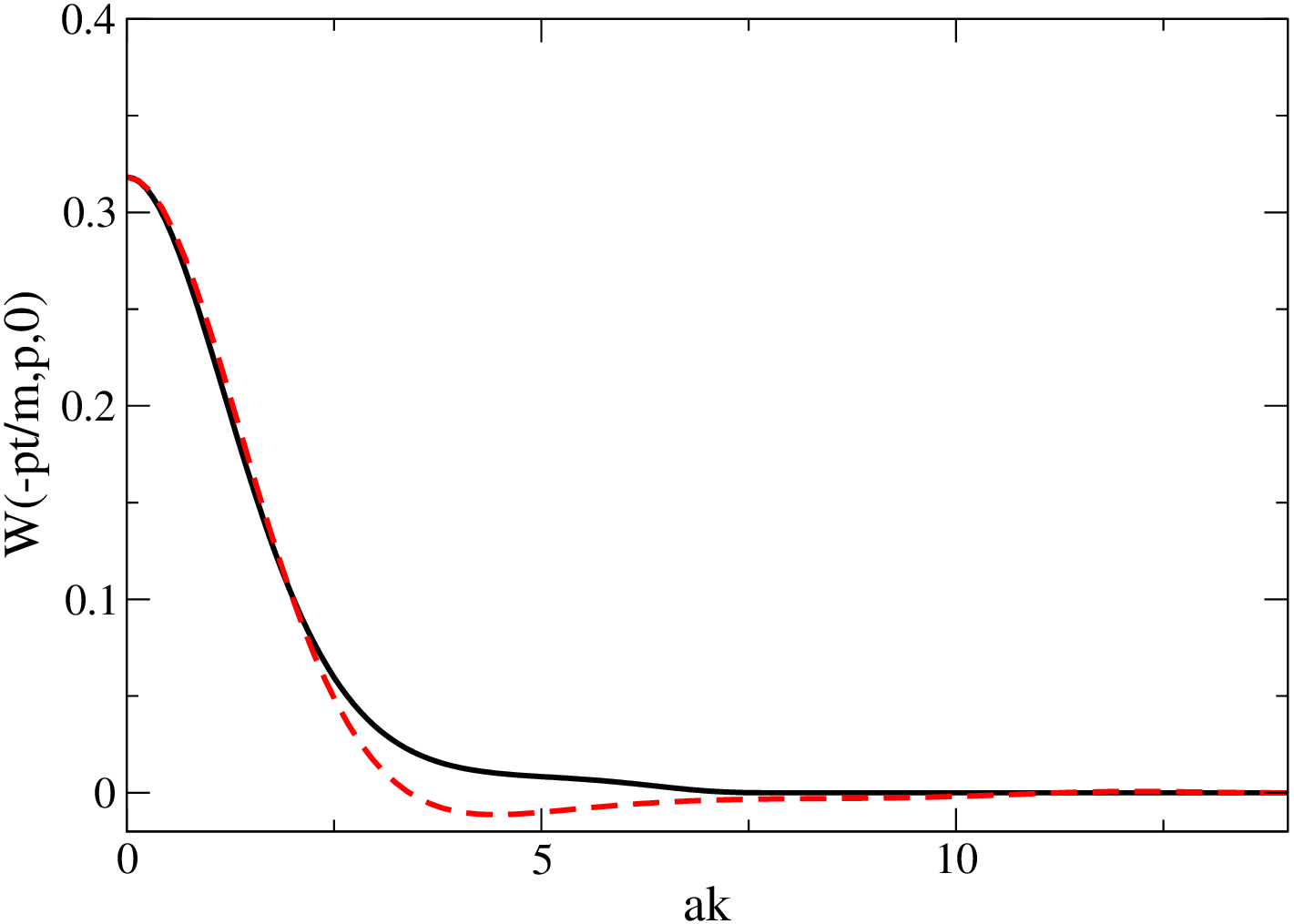}
\vspace{0.5cm}
\caption { Integrand $W(-\hbar kt/m,\hbar k,0)$
  for the calculation of $\rho_{qm}(0,t)$ using
  Eq. (\ref{Wigt}) for $t=0.14 t_0$ (full curve) and
  $t=0.07 t_0$ (dashed
  curve). }
\end{figure}

What we called the ``classical approximation'' can be given a different
interpretation. Comparing Eqs. (\ref{classical})
and (\ref{Wigt}) it can be viewed as the following
approximation for the Wigner function
\begin{equation}
  W(x,p,0)\approx |\langle x|\psi\rangle|^2    |\langle p|\psi\rangle|^2,
\end{equation}\\
i.e. a factorized probability distribution. As discussed above this
holds exactly only for a Gaussian wave packet.\\

It should be mentioned that various other wave packets have been studied
which are not groundstate wave functions of an initial potential,
e.g. a rectangular initial wave function \cite{Andrews,Schleich2}
which at short times leads to a strongly oscillatory behaviour. In 
reference 17 this is discussed also using the Wigner function.

\section {Conclusions}

The broadening of a free Gaussian wave packet is one of the few
time-dependent problems treated in courses on quantum mechanics.
The wave function $\langle x|\psi(t) \rangle$
can be calculated analytically to  obtain the
probability density  $|\langle x|\psi(t) \rangle|^2$.
It is little known that the time dependence of
other wave packets can differ considerably
from the smooth Gaussian broadening.

It is important to include the treatment of time-dependent Hamiltonians
when teaching quantum mechanics.
A simple case is the sudden change
from $\hat H_i$ to $\hat H_f$
treated in this paper.

It turned out that
directly adressing  $|\langle x|\psi(t) \rangle|^2$
using Eq. (3) allowed a simple way to obtain results
for harmonic oscillator systems. 
It is little known that the oscillatory behaviour in Fig. 1 can also
happen in the framework of classical statistical mechanics.
For the free time evolution after the quench
directly addressing  $|\langle x|\psi(t) \rangle|^2$ provides additional
insight, as the  introduction of
the concept of the Wigner function is straightforward.
This function  allows an understanding of the oscillatory behaviour
of the broadening of
 the initial ground state of an infinitely deep well
without performing a numerical
integration. This is an argument for presenting the concept
of the Wigner function in quantum mechanics courses.

\section{Acknowledgements}

The author would like to thank T. Dittrich, G. Hegerfeld,
F. Heidrich-Meisner, H. Leschke, S. Manmana and
V. Meden for a critical reading of the manuscript and useful suggestions.
Special thanks go to W. Schleich for bringing Refs. $8$ and
$17$ to my
attention and for a stimulating discussion about the use of Wigner functions.
 
\begin{appendix}

  \section{ The  free propagator}

In this appendix we shortly present the alternative derivation \cite{Andrews}
of the result for the quantum mechanical long time limit  Eq.(\ref{qmlt})
for the broadening of wave packets.

The time dependent wave function $\langle x|\psi(t)\rangle$ is calculated
by inserting position states
\begin{eqnarray}
\langle x |\psi(t)\rangle &=&  \int_{-\infty}^\infty \langle x|
e^{-i\hat Tt/\hbar}|x'\rangle\langle x'|\psi \rangle dx' \nonumber \\
&\equiv& \int_{-\infty}^\infty K(x.x',t)\psi(x')dx'~.
\end{eqnarray}
The propagator $K(x.x',t)$ can be calculated inserting momentum
eigenstates \cite{Merzb}
\begin{eqnarray}
K(x.x',t)&=& \int_{-\infty}^\infty \langle x|p\rangle e^{-i\epsilon_p t/\hbar}  
             \langle p |x'\rangle dx' \nonumber \\
 &=& \sqrt{\frac{m}{2\pi i \hbar t}}e^{im(x-x')^2/2\hbar t}~.
 \end{eqnarray}
 If one writes
 \begin{equation}
   \frac{m(x-x')^2}{2t\hbar}=\frac{mx^2}{2t\hbar}- \frac{mxx'}{t\hbar}
  +\frac{mx'^2}{2t\hbar} ~,
 \end{equation}
 the last term can be neglected in the long time limit for a sufficiently
 localized wave function $\psi(x')$
 and one obtains
 \begin{equation}   
|\psi(x,t)|^2  \to \frac{m}{t}|\tilde \psi(\frac{mx}{t})|^2 ~.
 \end{equation}  

\end{appendix}


\begin{thebibliography}{99}
\bibitem{CC} P. Calabrese and J. Cardy ``Quantum quenches in
  extended sysytems'', J. Stat. Mech. {\bf 2007}, P6008, 1-33
  \bibitem{Huang} K. Huang, {\it Introduction to Statistical Physics}
 (Taylor and Francis, London, 2001)
  \bibitem{Wigner} E. Wigner, ``On the quantum correction for thermodynamic
    equlibrium'', Phys. Rev. {\bf 40}, 749-759 (1932)
\bibitem{Case} W.B. Case, ``Wigner functions and Weyl transforms for
  pedestrians'', Am. J. Phys. {\bf 76}, 937-946 (2008)
\bibitem{Merzb} E. Merzbacher, {\it Quantum Mechanics}, 3rd edition
  (John Wiley and Sons, New York, 2000)
\bibitem{Ball} L.E. Ballentine, {\it Quantum Mechanics: A Modern
    Development}, 2nd edition (World Scientific Publishing)
\bibitem{Andrews} M. Andrews,
      ``The evolution of free wave packets'' ' Am. J. Phys. {\bf 76},
      1102-1107 (2008)
\bibitem{Schleich1} K. Vogel, F. Gleisberg, N.L. Harshman, P. Kazemi,
  R. Mack, L. Plimak, W.P. Schleich
  ``Optimally focusing wave packets'',
  Chemical Physics, {\bf 375}, 133-143 (2010)
\bibitem{Holstein} B.R. Holstein, ``Topics In Advanced Quantum Mechanics'',
      (Adison-Wesley, Redwood City, 1992), p. 108-110
 \bibitem{Nauenberg} M. Nauenberg, ``Einstein's equivalence principle
  in quantum mechanic revisited'', Am. J. Phys. {\bf 84}, 879-882 (2016)
\bibitem{Jim} A.M. Orjuela and J.K. Freericks, ``Free expansion of a
  Gaussian wavepacket using operator manipulations'',
  Am. J. Phys. {\bf 91}, 463 (2023)
\bibitem {Schleich} W. P. Schleich, {\it Quantum optics in Phase Space},
(John Wiley and Sons, New York, 2001)
\bibitem {WFST} D. K. Ferry and M. Nedjalkov, {\it The Wigner function
    in Science and Technology} (IOP Publishing, 2018)
\bibitem {Hudson} R.L. Hudson, ``When is the Wigner quasi-probability density
  non-negative?'', Reports on Mathematical Physics, {\bf 6}, 249-252
  (1974)
\bibitem{Mita} K. Mita
      ``Dispersion of non-Gaussian free particle wave packets''
   Am. J. Phys. {\bf 76}, 950-953 (2007)  
  \bibitem{Thomas} T. Dittrich, {\it INFORMATION DYNAMICS In Classical
    and Quantum Systems} (Springer Nature Switzerland, 2022) , p. 276
\bibitem{Schleich2} M.R. Goncalves, W.B. Case, A. Arie, W.P. Schleich,
  ``Single-slit focusing and its representations'', Appl. Phys. B,
  123:121 (2017)

  
\end{thebibliography}
\end{document}